\documentclass[aps,amsmath,nofootinbib,superscriptaddress,amssymb,preprintnumbers,floatfix,showpacs,preprint]{revtex4-1}
 \usepackage{amsmath,txfonts,longtable,booktabs,overpic,amssymb,bm,bbm,multirow,float,graphicx,color,dcolumn,subfigure,hyperref,tikz}
\definecolor{blue}{RGB}{45,48,146}
 \hypersetup{colorlinks,citecolor=black,anchorcolor=black,menucolor=black, linkcolor=black,filecolor=black,runcolor=black,urlcolor=black,frenchlinks=true}

\begin{document}

	\title{The enhancement of pair production in oscillated overlapped fields}

  \author{Adiljan Sawut}
	\affiliation{ State Key Laboratory for Tunnel Engineering, China University of Mining and Technology, Beijing 100083, China}
	
	\author{Ying-Jun Li}\email{lyj@aphy.iphy.ac.cn}
	\affiliation{ State Key Laboratory for Tunnel Engineering, China University of Mining and Technology, Beijing 100083, China}

\author{Miao Jiang}
\affiliation{School of Science, China University of Mining and Technology, Beijing 100083, China}

 \author{Bai-Song Xie}\email{bsxie@bnu.edu.cn}
	\affiliation{ Key Laboratory of Beam Technology of the Ministry of Education, and College of Nuclear Science and Technology, Beijing Normal University, Beijing 100875, China}
\affiliation{Institute of Radiation Technology, Beijing Academy of Science and Technology, Beijing 100875, China}

\begin{abstract}
  The influence of potential well width on electron-positron pair production has been examined through theoretical and numerical approaches by employing the computational quantum field theory. Quantum interference effects in pair production is investigated in the two overlapped potential wells with varied widths and frequencies. Several dominant processes, involving the absorption of an integer number of photons, significantly impact on pair production. Notably, specific multiphoton absorption processes exhibit distinct changes as the potential well width expands, with the absorption of four photons process displaying noteworthy effects. Additionally, the influence of the smaller frequency to the yield of the pair production can not be ignored and the most optimized frequencies in our overlapped fields has been studied and exhibited.
\end{abstract}
\maketitle

\section{Introduction\label{sec:1}}
The laser is the most powerful tool to study the strong field quantum electrodynamics (QED) in laboratory frame \cite{Burke:1997ew} and had enormous breakthrough since the invention of the Chirped Pulsed Amplification (CPA) technics \cite{Strickland:1985gxr,MacLeod:2022qid}. Researchers were optimistic \cite{Mourou:2006zz,Chen:2002nd} to the development of the laser technology that they had estimated the intensity of the laser would have a rapid growth and approaches Schwinger critical field soon\cite{Schwinger:1951nm}. However, the current laser intensity is five to six orders of magnitude smaller, which is $10^{23}\sim10^{24}\rm{W/cm^2}$ \cite{ELI,XCELS}. Meanwhile, many theoretical works have been performed, and one of them is the electron-positron pair production that is the topic studied in this paper.

The study of electron-positron pair production \cite{Sauter:1931zz,Affleck:1981bma,Kim:2000un,Dunne:2005sx} in intense electromagnetic fields is a rich area of theoretical physics, where a variety of methods have been developed to tackle this nonlinear QED process \cite{Su:2019nwr,Gong:2018zqs}. The Wentzel-Kramers-Brillouin (WKB) approximation is a semi-classical approach that allows for an analytical understanding of tunneling processes, which are essential for describing pair production in strong fields \cite{DiPiazza:2004lsj,Dumlu:2010ua,Dumlu:2011rr,Strobel:2014tha}. The Dirac-Heisenberg-Wigner (DHW) formalism provides a phase-space description of quantum fields, offering a comprehensive framework to study the real-time dynamics of pair production \cite{Hebenstreit:2011wk,Li:2015cea,Li:2017qwd,Olugh:2018seh,Hu:2023pmz,Xie:2017xoj}. The quantum Vlasov equation (QVE), a kinetic equation for the distribution function of particles in phase space, has been adapted to include QED effects, enabling the study of plasma oscillations and instabilities that can occur in the presence of strong fields \cite{Kluger:1991ib,Alkofer:2001ik,Hebenstreit:2009km,Nuriman:2012hn,Li:2014psw,Kohlfurst:2013ura,Gong:2019sbw}. Computational quantum field theory (CQFT) \cite{Braun:1999zza,Krekora:2004trv} has been very helpful in providing numerical solutions to scenarios where analytical methods fall short, allowing for the exploration of pair production in various field configurations \cite{Jiang:2013wha,Su2012MagneticCO,Tang:2013qna,Sawut:2021cjr,Lv:2018sqm,Su:2019jot,Zhou:2019euj}. These theoretical tools each offer unique insights into the complex nature of pair production, and their development and application continue to enhance our understanding of quantum phenomena.

In this paper, we study the pair production process in two overlapped oscillating laser fields with different frequencies  and varied width of the potential well by using the CQFT method. As the width between two fields gets wider the interference effects can be observed and emerge a periodic variation in both pair production probability and production rate. These phenomena can be explained by the first-level perturbation theory conveniently \cite{Jiang:2013wha}. Multiphoton effects can be seen from the transition energy probability distribution $U_{pn}$ picture specifically, each process has been influenced by the width of the potential well and some processes play main role in this scenario.

Our paper is organized as follows. In Sec.~\ref{sec:2}, we give a brief introduction to our theoretical framework that allow us to calculate the energy and momentum distribution, number of created pairs in 1+1 dimensional fields. In Sec.~\ref{sec:3}, we provide the theoretical results and explanations of studying the different features of pair production by considering the variable width of the potential well. The Sec.~\ref{sec:4} is a summary for our work.

We use the atomic units ($\hbar=m_e=e=1$) throughout this paper.

\section{Theoretical framework and the background field}\label{sec:2}

CQFT is a method that numerically solves the Dirac equation in all time and space \cite{Wang:2020wyp,Wu:2019weg,Zhou:2021hko,Zhou:2019euj}. Thus, the calculation begins with the Dirac equation,
\begin{equation}
i\partial\hat{\psi}\left({z},t\right)/\partial{t}=\lbrack{c\alpha_z{\hat{p}}_z+\beta c^2+V(z,t)}\rbrack\hat{\psi}\left(z,t\right),
\end{equation}
where $\alpha_{z}$ and $\beta$ denote the Dirac matrices, and we will focus on a single spin due to the lack of magnetic field in our one-dimensional system. As a results, the four-component spinor wave function becomes two components, and the Dirac matrix $\alpha_{z}$ and $\beta$ are replaced by the Pauli matrices $\sigma_1$ and $\sigma_3$ \cite{Tang:2013qna}.
The field operator $\hat{\psi}\left({x},t\right)$ can be expanded in both time-depended and time-independed complete bases as follows:
\begin{equation}
\hat{\psi}\left({x},t\right)=\sum_{{p}}{\hat{b}}_{p}(t)u_{p}({x})+\sum_{n}\ {\hat{d}}_{n}^\dag(t) u_{n}({x})
\end{equation}
\begin{align}\label{eq:2}
= \sum_{{p}}{\hat{b}}_{p}u_{p}({x,t})+\sum_{n}\  {\hat{d}}_{n}^\dag u_{n}({x,t}),
\end{align}
where $u_{p}({x})$ and $u_{n}({x})$ are the field-free energy eigenstates, $u_{p}({x,t})$ and $u_{p}({x,t})$ are the time evolution of the $u_{p}({z})$ and $u_{n}({z})$ or can be explained as time-dependent wave function. ${\hat{b}}_{p}$ and ${\hat{d}}_{n}^\dag$ are the electron annihilation and the positron creation operators, the time-dependent creation and annihilation operators can be written as follows:
\begin{equation}
  {\hat{b}}_p(t)=\sum_{{p^\prime}}{{\hat{b}}_{p^\prime}\int dx u_{p}^{*}({x})u_{p^\prime}\left({x},t\right)+\sum_{n^\prime}\ {\hat{d}}_{n^\prime}^\dag \int dx u_{p}^{*}({x})u_{n^\prime}\left({x},t\right),\ \ \ \ \ \ \ \ \ \ }
  \end{equation}
  \begin{equation}
  {\hat{d}}_n^\dag(t)=\sum_{{p^\prime}}{{\hat{b}}_{p^\prime}\int dx u_{n}^{*}({x})u_{p^\prime}\left({x},t\right)+\sum_{n^\prime}\ {\hat{d}}_{n^\prime}^\dag \int dx u_{n}^{*}({x})u_{n^\prime}\left({x},t\right).\ \ \ \ \ \ \ \ \ \ }
  \end{equation}

For the spatial distribution of electrons, it is defined by the positive energy part of the field operator as $\hat{\psi}_{e}\left({x},t\right)\equiv\sum_{{p}}{\hat{b}}_{p}(t)u_{p}({x})$, and the spatial particle number density of created electrons can be written as
\begin{equation}
  \rho_{e}\left({x},t\right)=\langle\langle\text{vac}\mid\mid {\hat{\psi}}_{e}^{+\dag}\left({x},t\right){\hat{\psi}}_{e}^+\left({x},t\right)\mid\mid\text{vac}\rangle\rangle
  \end{equation}
  \begin{equation}
    =\sum_{{n}}{\mid\sum_{{p}}{U_{pn}\left(t\right)u_p\left({x}\right)\mid^2,\ }}
    \label{density}
    \end{equation}
where $U_{p,n}\left(t\right)=\langle u_{p}(x)\mid U(t)\mid u_{n}(x)\rangle$ and $U(t)=\hat{T} {\rm exp}(-i \int H dt)$ represents the time evolution operator. Energy density of the electrons are expressed as
\begin{equation}
\rho({E_p},t)=\sum_{{n}}{\mid{U_{p,n}\left(t\right)\mid^2,}}
\end{equation}
and Eq.~\eqref{density} integral over the space to obtain the total number of created electrons as follow:
\begin{equation}
  N(t)=\int dx\rho\left({x},t\right)=\sum_{p,n}{\mid{U_{p,n}\left(t\right)\mid^2.}}
  \end{equation}

  Our theoretical frameworks are fully ready for the calculation to the next step and insert the background field to study the pair production processes.

\section{Numerical results}\label{sec:3}
\begin{figure}[ht!]\centering
  \includegraphics[width=0.7\textwidth]{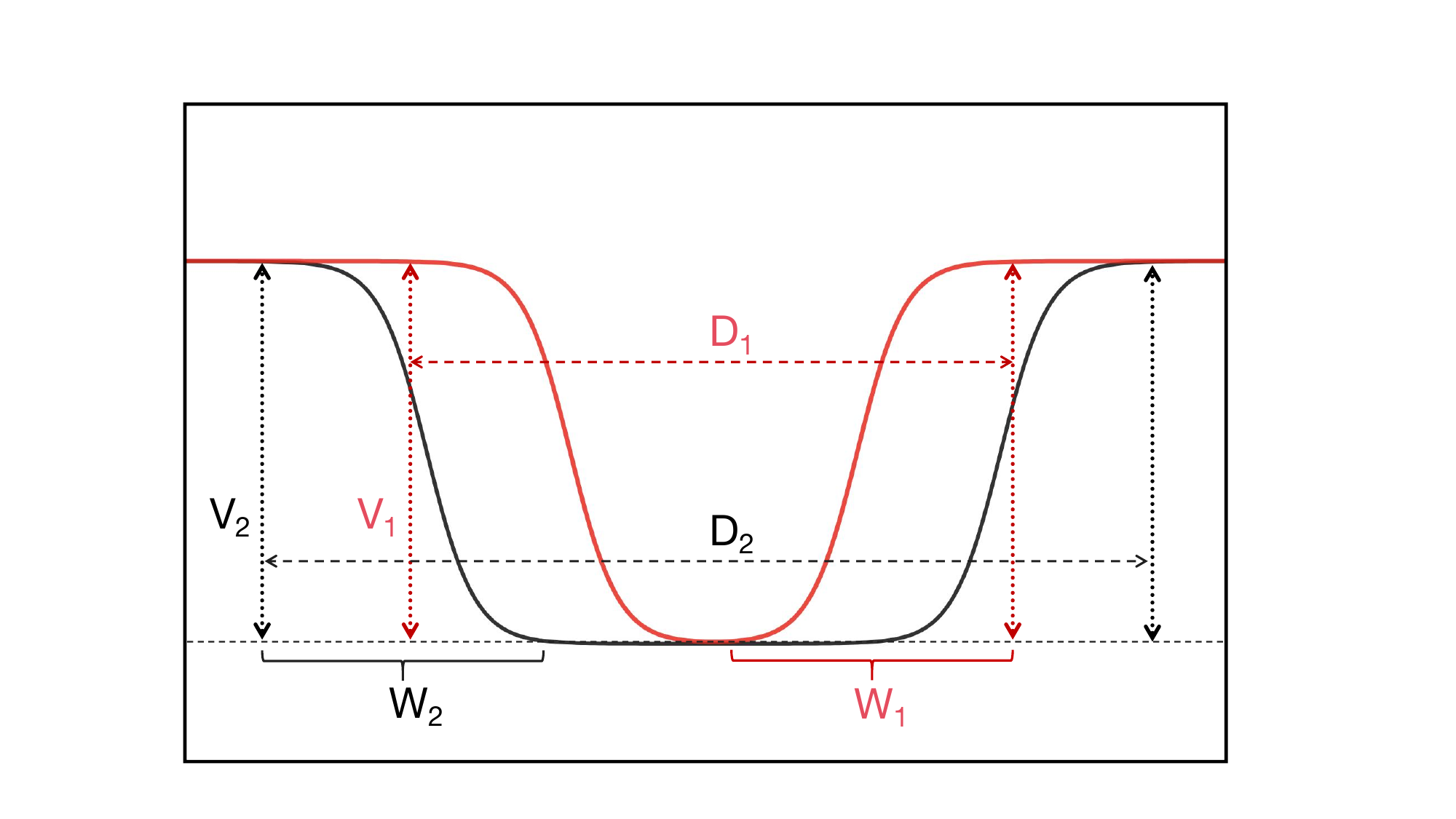}
  \caption{Shape of the two overlapped potential wells where $D_1$ and $D_2$ are for the widths of the potential well, $W_1$ and $W_2$ are the widths of the electric field, $V_1$ and $V_2$ are the height the well.
  \label{fig:1}}
  \end{figure}

In this work our field configuration for the potential well can be written as follow:
 \begin{equation}
    V(z)=\frac{V_1}{2} S_1(z){\rm sin}({\omega}_1{t})+\frac{V_2}{2}S_2(z){\rm sin}({\omega}_2{t}),
    \label{eqV}
    \end{equation}
    where $S_1(z)$ and $S_2(z)$ are the shape of our Sauter field and can be expressed as $S_1(z)={\rm tanh}((x-D_1/2)/W_1)+{\rm tanh}((x+D_1/2)/W_1)$ and $S_2(z)={\rm tanh}((x-D_2/2)/W_2)+{\rm tanh}((x+D_2/2)/W_2)$, $V_1$ and $V_2$ are the depth of the each well, $D_1$ and $D_2$ are the widths of the potential wells, $\omega_1$ and $\omega_2$ are the frequencies of the fields, $W_1$ and $W_2$ are the widths of each electric field and we will use $W$ for both of them due to $W_1=W_2$. The shape of the potential well can be visualized as Fig.~\ref{fig:1}.

It is shown in Eq.~\eqref{eqV} that the two colored fields will act on the whole process to trigger pair production via multiphoton effect. We set the correlated parameters as, $V_1$=$V_2=2c^2$, $\omega_1=0.7c^2$ and $\omega_2=1.7c^2$, $D_1$ and $D_2$ can be varied from $1/c$ to $20/c$ and $W=2/c$, where $c$ is the speed of light ($c=137.036$ a.u).

\begin{figure*}[!htb]
      \includegraphics[width=0.49\linewidth]{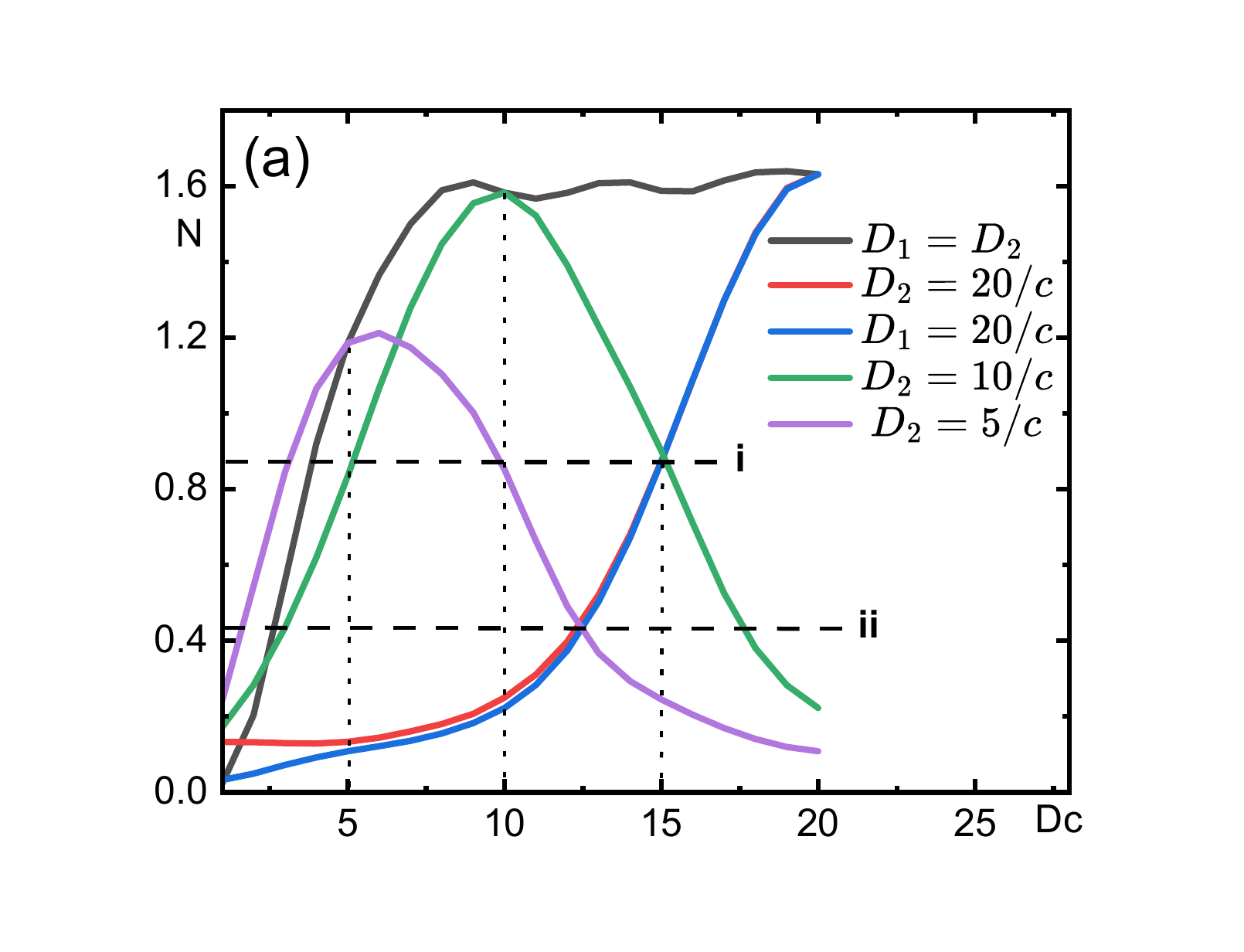}
      \includegraphics[width=0.49\linewidth]{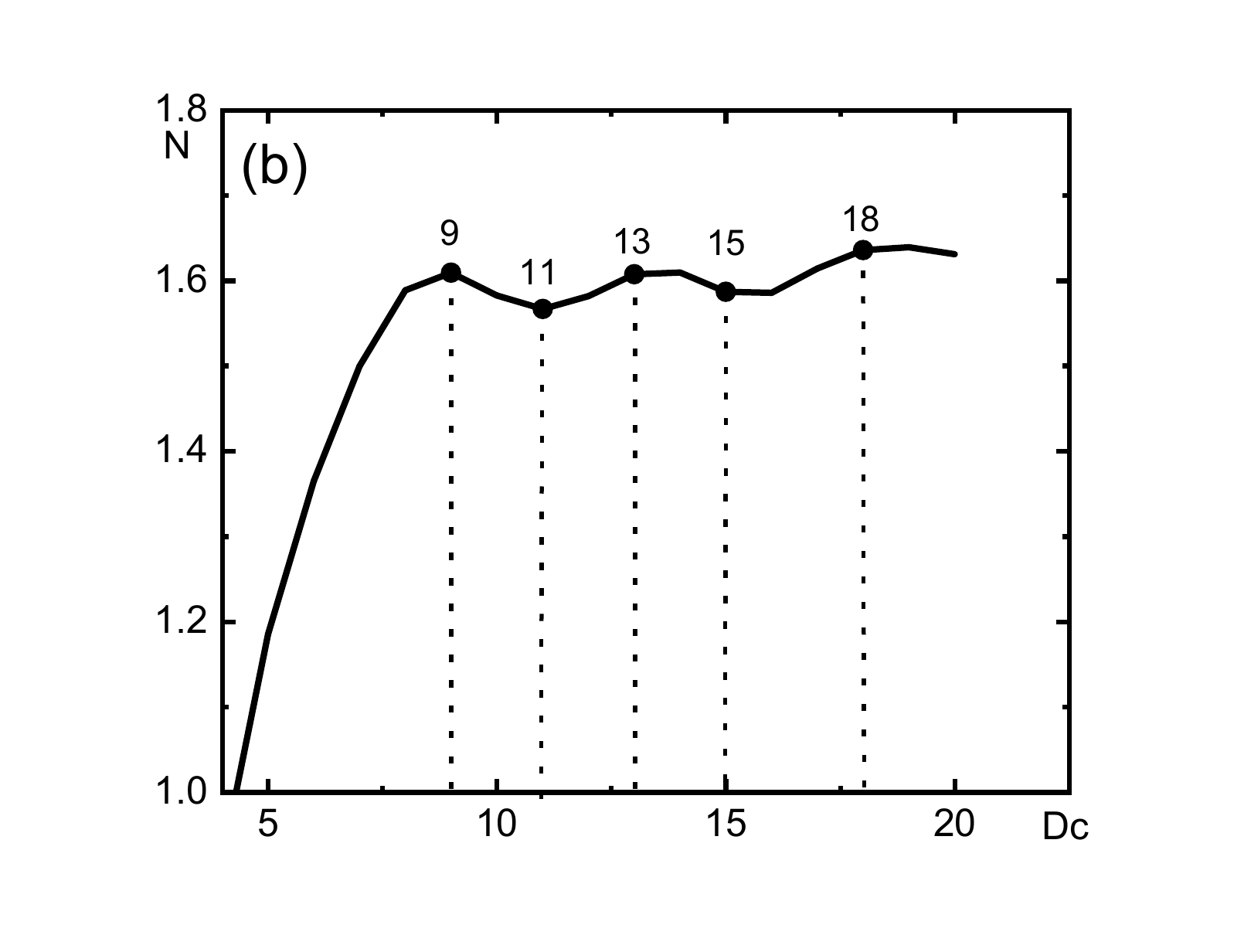}
      \caption{Yield of the pair production with different $D_1$ and $D_2$ sets, for (a) $D_1=D_2$ varies from $1/c$ to $20/c$ (black line), $D_1=20/c$ and $D_2$ varies from $1/c$ to $20/c$ (blue line), other three sets $D_2=20/c, 10/c, 5/c$ separately, and $D_1$ varies from $1/c$ to $20/c$. For (b) exactly the same result for black line in Fig.~\ref{Fig:Data1}(a) for highlighting. Other parameters used in the simulations are $t=0.003$, $\omega_1=0.7c^2$ and $\omega_2=1.7c^2$, $V_1=V_2=2c^2$ and $W=2/c$.}\label{Fig:Data1}
 \end{figure*}

  After numerically solving the Dirac equation by using CQFT method in overlapped potential wells, we have shown the yield of the pair production in time $T=0.003$ (a.u) by changing the width of the potential well in five different sets. First set is where $D_1$ and $D_2$ vary from $1/c$ to $20/c$ at the same time, as shown in Fig.~\ref{Fig:Data1}(a) (black line). Second set is $D_2$ varies from $1/c$ to $20/c$ and $D_1=20/c$ (blue line) and other three sets are the variation of $D_1$ from $1/c$ to $20/c$ where $D_2$ takes different values, $D_2=20/c, 10/c, 5/c$, separately.

  As shown in Fig.~\ref{Fig:Data1}(a), for the first set, the yield of the pair production increases linearly from $1/c$ to $9/c$ and starts to change periodically until the last value of $D$, the periodic variation is much more noticeable where the width $W$ gets smaller, see the reference \cite{Jiang:2013wha}. Such a change in the yield of creation with different potential width $D$ is a coherence effect, which is due to the interference of the created particles from both localized fields. The yields of pair production for second set(blue line) and third set(red line) are indistinctive after $5/c$. However, we can be able to discover there are slight dominant position for the third set in the yield of pair production where $D_2=20/c$ and $D_1$ varies from $1/c$ to $20/c$. It can be explained by the dominant frequency of $\omega_2=1.7c^2$ still has the advantage of creating pairs and the smaller frequency $\omega_1=0.7c^2$ also plays an important role as a dynamically assisted field. For the potential width with $D_2=10/c$ and $5/c$, they both increase and reach the highest value at $D_1=D_2$, then start to decrease until $D_2=20/c$. As for the production yield reaches the highest value at $D_1=D_2$, the two oscillated field overlapped exactly one on the other and increase the intensity of the combined field and starts to trigger the creation of more pairs.

  There is another interesting phenomenon that the yield of pair production has the same value for $D_1=15/c, D_2=20/c$ and $D_1=15/c, D_2=10/c$, we can easily observe the crossed value of blue line and green line from Fig.~\ref{Fig:Data1}(a). To explain this phenomenon we have drawn two isolines (i) and (ii) in Fig.~\ref{Fig:Data1}(a) and discover the yields of pair production are exactly same for $D_1=5/c, D_2=10/c$ and $D_1=15/c, D_2=10/c$ in green line, $D_1=15/c, D_2=20/c$ in red line(same value with blue line) and $D_1=10/c, D_2=5/c$ for purple line. Our assumption is the yield of pair production are the same for different overlapped fields when the absolute value of $|D_1-D_2|\geq5/c$. So we have demostrated the isoline (ii) to emphasize the validity of our assumption. As for isoline (ii), the yields of pair production are in same values for $D_1=3/c, D_2=10/c$ and $D_1=17/c, D_2=10/c$ in green line, $D_1=12/c, D_2=5/c$ in purple line and $D_1=13/c, D_2=20/c$ in red line(same value with blue line), where $|D_1-D_2|=7/c$.

\begin{figure*}[!htb]
  \includegraphics[width=0.49\linewidth]{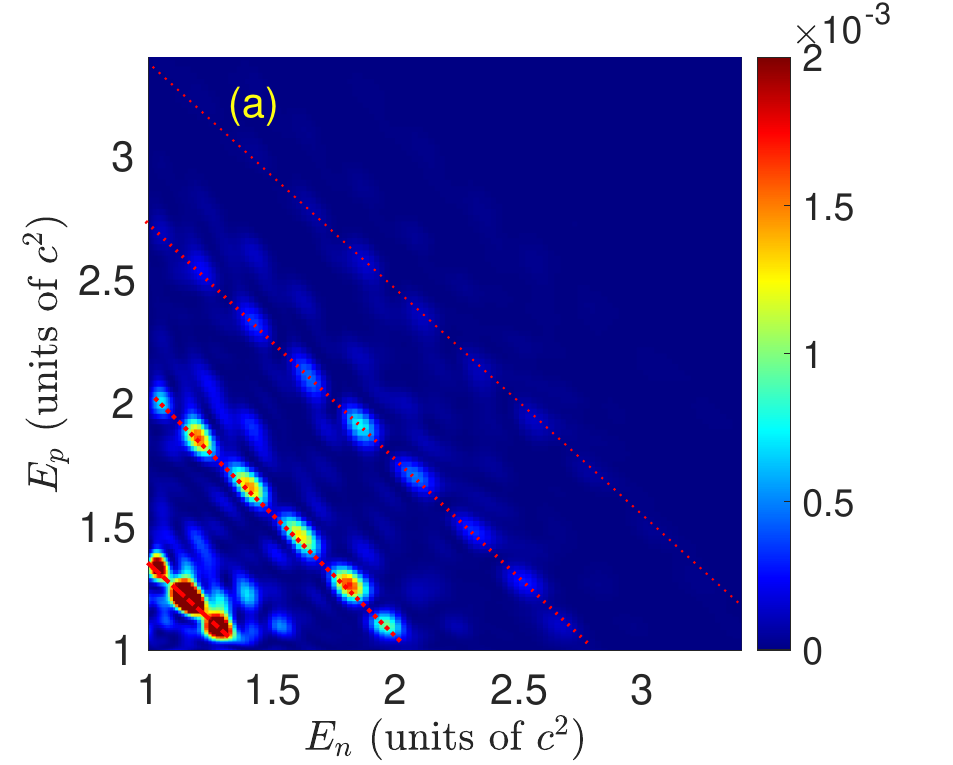}
  \includegraphics[width=0.49\linewidth]{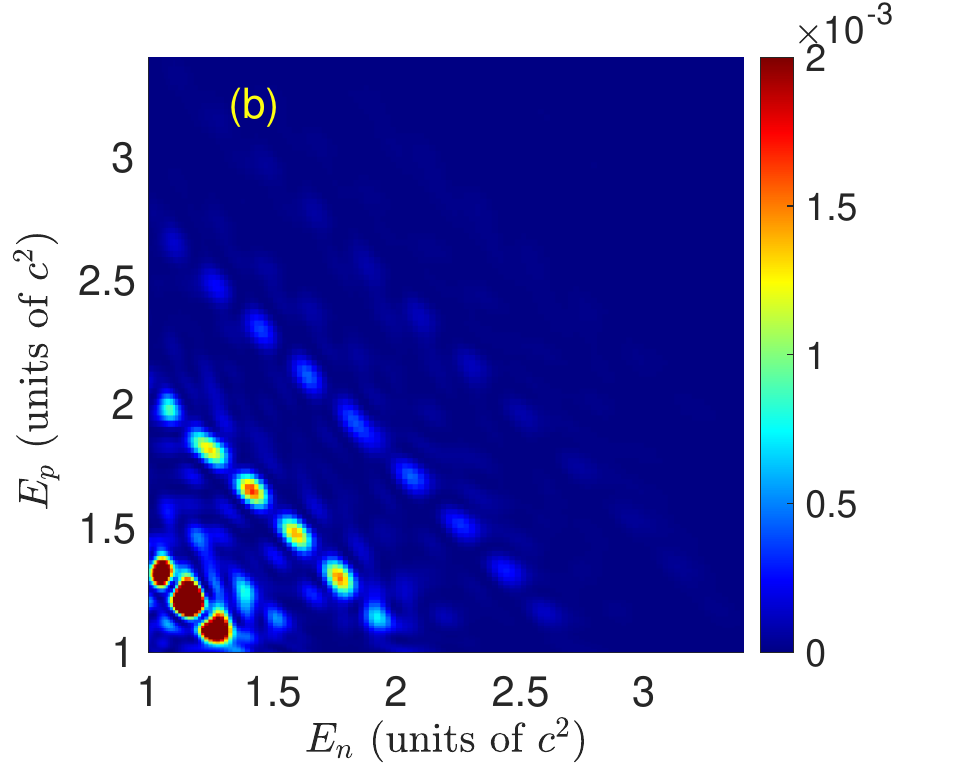}
  \includegraphics[width=0.49\linewidth]{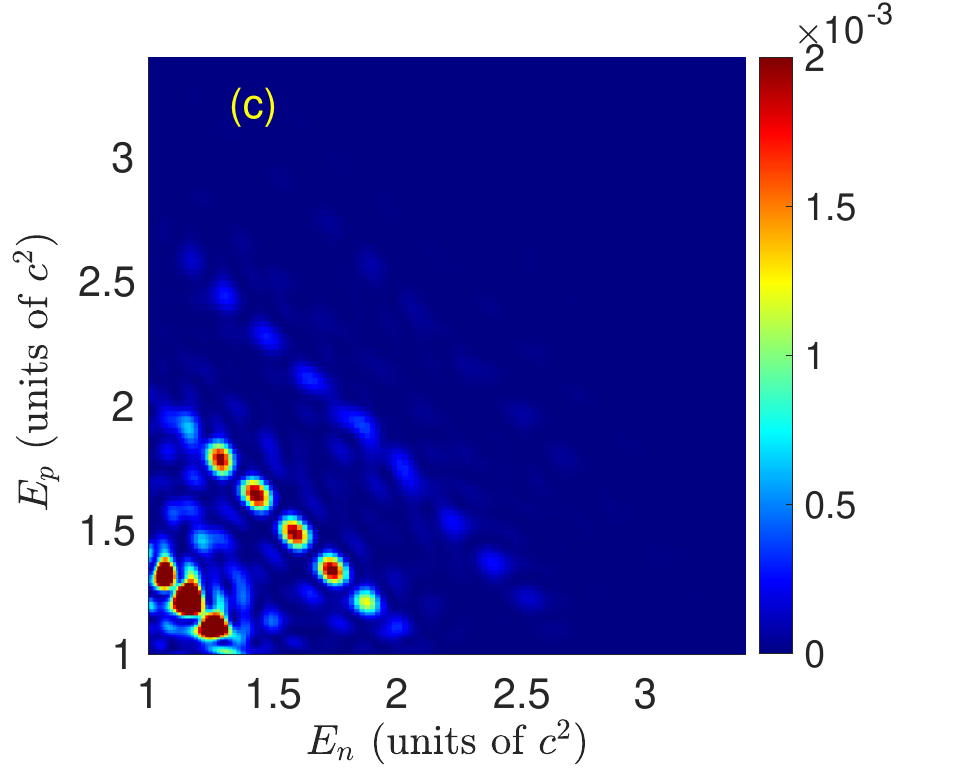}
  \includegraphics[width=0.49\linewidth]{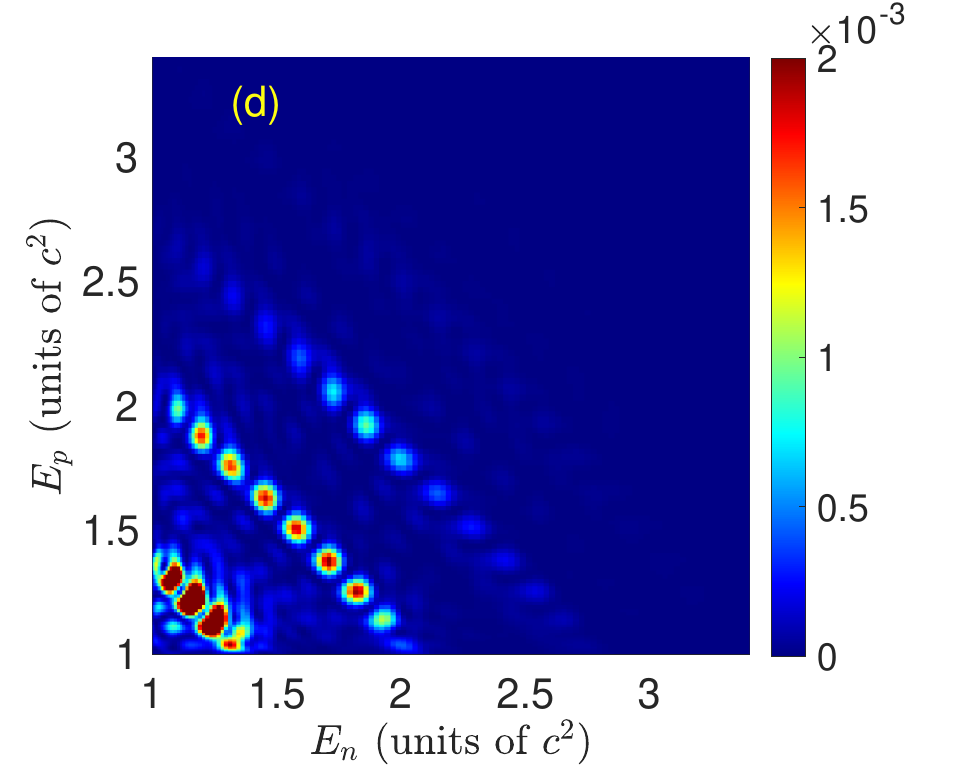}
  \caption{Contour plot of probability distribution of created pairs transition energy in oscillating fields with $\omega_1=0.7c^2 , \omega_2=1.7c^2$ ,$V_1=V_2=2c^2$ and the width $W=2/c$. From (a) to (d), corresponding width of the potential well are $D=11/c, 13/c, 15/c, 18/c$. Other parameters are same with the Fig.~\ref{Fig:Data1}.
  \label{Fig:Data2}}
\end{figure*}

As for the periodic changes in the yield, we have investigated four different widths of potential well in Fig.~\ref{Fig:Data1}(b), namely $D_1=D_2=11/c, 13/c, 15/c, 18/c$ and present the corresponding transition energy distribution probability $U_{pn}$ in Fig.~\ref{Fig:Data2}. The horizontal and vertical axis represent the positive and negative energy states, respectively. Each intersection point of the axis corresponds to a specific negative energy state, indicating the probability of transition to a certain positive energy state. The color at each point represents the probability of transition when the absorbed energy is the sum of the horizontal and vertical axis, $E_p + E_n$. As for Fig.~\ref{Fig:Data2}(a), it is the $U_{pn}$ contour graph for $D_1=D_2=11/c$, (b) for $D_1=D_2=13/c$, (c) for $D_1=D_2=15/c$ and (d) for $D_1=D_2=18/c$. The difference between those four figures is observable.

  To provide a more precise explanation, we use Fig.~\ref{Fig:Data2}(a) to highlight the main processes in red, which are separately corresponded to the processes $\omega_1+\omega_2=2.4c^2, 2\omega_1+\omega_2=3.1c^2, 3\omega_1+\omega_2=3.8c^2, 4\omega_1+\omega_2=4.5c^2$ from left to the right. The most dominant process in here is $\omega_1+\omega_2$ due to the total energy of two photons are larger than the threshold energy  the pair creation, $\omega_1+\omega_2=2.4c^2>2c^2$, pairs are created continuously at a constant rate. Obviously, the probability distribution of this process in Fig.~\ref{Fig:Data2}(a) is smaller than the distribution in Fig.~\ref{Fig:Data2}(b). For the second major process which is $2\omega_1+\omega_2=3.1c^2$, has indistinctive changes from Fig.~\ref{Fig:Data2}(a) to Fig.~\ref{Fig:Data2}(b). However, for the third process $3\omega_1+\omega_2=3.8c^2$, only in Fig.~\ref{Fig:Data2}(b) and Fig.~\ref{Fig:Data2}(d) is more noticeable and these exactly correspond to the higher creation yield in Fig.~\ref{Fig:Data1} with the width of $D=13/c$ and $D=18/c$. The fourth process $4\omega_1+\omega_2=4.5c^2$ has the lowest probability distribution and hardly noticeable in Fig.~\ref{Fig:Data2}(a) and Fig.~\ref{Fig:Data2}(c). Although each processes are influenced by the different potential width, but the third process $3\omega_1+\omega_2=3.8c^2$ has more distinctive variation in the yield of pair production.

  Another interesting phenomenon is that the probability distribution of transition energy has started to disperse into many small pieces as the potential width gets larger, see the first and second major process $\omega_1+\omega_2=2.4c^2$, $2\omega_1+\omega_2=3.1c^2$. For example, the second major process changes from 6 small areas in Fig.~\ref{Fig:Data2}(a) to 9 small areas in Fig.~\ref{Fig:Data2}(d).

  \begin{figure}[ht!]\centering
    \includegraphics[width=1\textwidth]{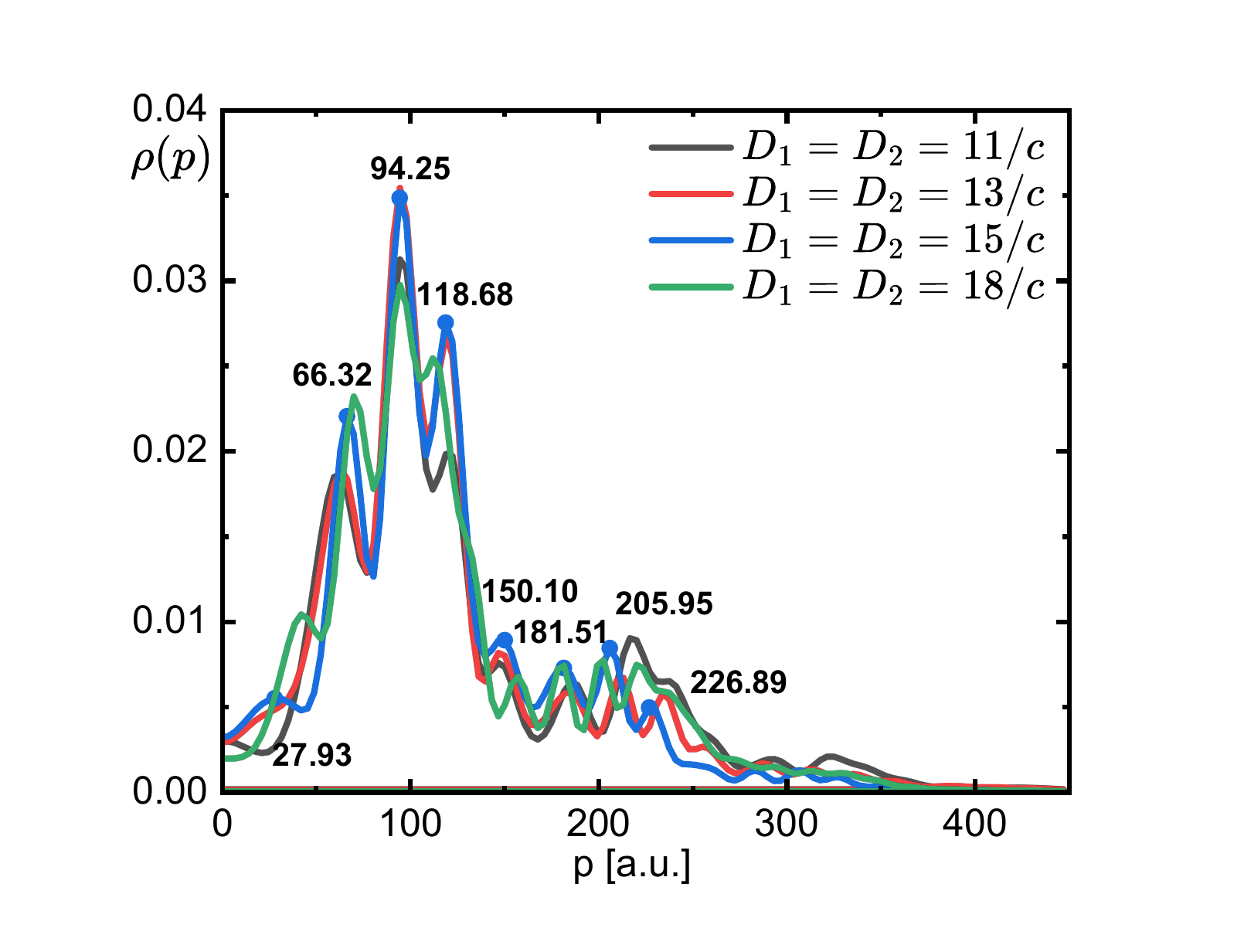}
    \caption{Momentum spectrum of the created pairs at $t=0.003$ with parameters of $V_1=V_2=2c^2$, $\omega_1=0.7c^2$ and $\omega_2=1.7c^2$, $W=2/c$, for $D_1=D_2=11/c, 13/c, 15/c, 18/c$. The given numbers are the peak values for $D_1=D_2=18/c$.
    \label{fig:4}}
    \end{figure}

For the further demonstration of periodically variation in the yield, we also share the results for the momentum spectrum for different potential well widths in Fig.~\ref{fig:4}. For this four different sets, we are aware of that $D_1=D_2=13/c$ and $D_1=D_2=18/c$ have the higher value of pair production yield than $D_1=D_2=11/c$ and $D_1=D_2=15/c$ from the Fig.~\ref{Fig:Data1}(b). The momentum spectrum for different $D$ likely have no distinctive advantages, however, we can observe very slight optimization for $D_1=D_2=13/c$ (red line) and $D_1=D_2=18/c$ (green line) in Fig.~\ref{fig:4} around lowest momentum range.

  As we have discussed above in detail, periodically variation in the yield and the dispersion of the probability distribution of the transition energy are the most interesting phenomena. Those can be explained by deriving the first-order perturbation theory, as in reference \cite{Jiang:2013wha}. The first-order transition amplitude from negative state $\mid{n}\rangle$ to the positive state $\mid{p}\rangle$ at arbitrary time $t$ is given by:
  \begin{equation}
    C^{(1)}_{pn}=\frac{1}{i}\int_0 ^t \langle p \mid H^\prime \mid n \rangle e^{i(E_p-E_n)\tau} d\tau.
    \end{equation}
  After we insert the interaction part of the Hamiltonian, we obtain the expression:
  \begin{equation}
    C^{(1)}_{pn}=\pi W V_0 \frac{A_{pn}}{2} {\rm csch} [\pi W \frac{(p+n)}{2}] \cos[\frac{(p+n)D}{2}]f(t),
    \label{Cpn}
  \end{equation}
  where $f(t)$ is a time depended function related to the field frequency $\omega$ and due to our study object is width $D$, the $f(t)$ is a time-depended part which is unnecessary to be written specifically. The inner product $A_{pn}$ can be defined as:
  \begin{equation}
    A_{pn}=\frac{{\rm sgn}(n)\sqrt{E_p+c^2}\sqrt{-E_n-c^2}+ {\rm sgn}(p)\sqrt{E_p-c^2}\sqrt{-E_n+c^2}}{4\pi \sqrt{-E_pE_n}}.
  \end{equation}

  As shown in Eq.~\eqref{Cpn}, periodically changes for the probability distribution of the transition energy comes from the $\cos[(p+n)D/2]$ part. By summing over all the states of $p$ and $n$, we can obtain the first-order perturbation estimation of the total pair production as:
  \begin{equation}
    N^{(1)}=\sum_{pn}\mid{C^{(1)}_{pn}}\mid ^2,
    \label{N}
  \end{equation}
the potential width $D$ related term in last expression $\cos^2[(p+n)D/2]$ leads to the periodical variation in yield of pair production with the changes of the potential width $D$, the interference effect is also more pronounced. Using equation $\cos^2[(p+n)D/2]$ in perturbation theory we can calculate the oscillation periods for yields with momentum $p$ in Fig.~\ref{fig:4} for different width of potential well. As for $D_1=D_2=15/c$, we can easily calculate the oscillation period as $\Delta p=28.69$ and $p=\sqrt{({E_p}^2-c^4)/c^2}=90.9$, which is the highest value of the momentum. We can list the other peak values for momentum perturbatively and compare the results with the peak values in Fig.~\ref{fig:4}, there has an acceptable margin of error, for example, the momentum we have calculated $p=90.9$ and the highest peak value $p=94.25$ shown in Fig.~\ref{fig:4}, the peak value of $p=118.68$ and the $p+\Delta p=90.9+28.69=119.59$.

As we have mentioned above, there is a slight advance in the yield for red line in Fig.~\ref{Fig:Data1} where $D_1$ keeps varying and $D_2=20/c$ than the blue line where $D_1=20/c$ and $D_2$ varies. Our assumption is the yield of pair production is also sensitive to the small frequency as a dynamically assisted to the electron-positron production process.
\begin{figure*}[!htb]
  \includegraphics[width=0.49\linewidth]{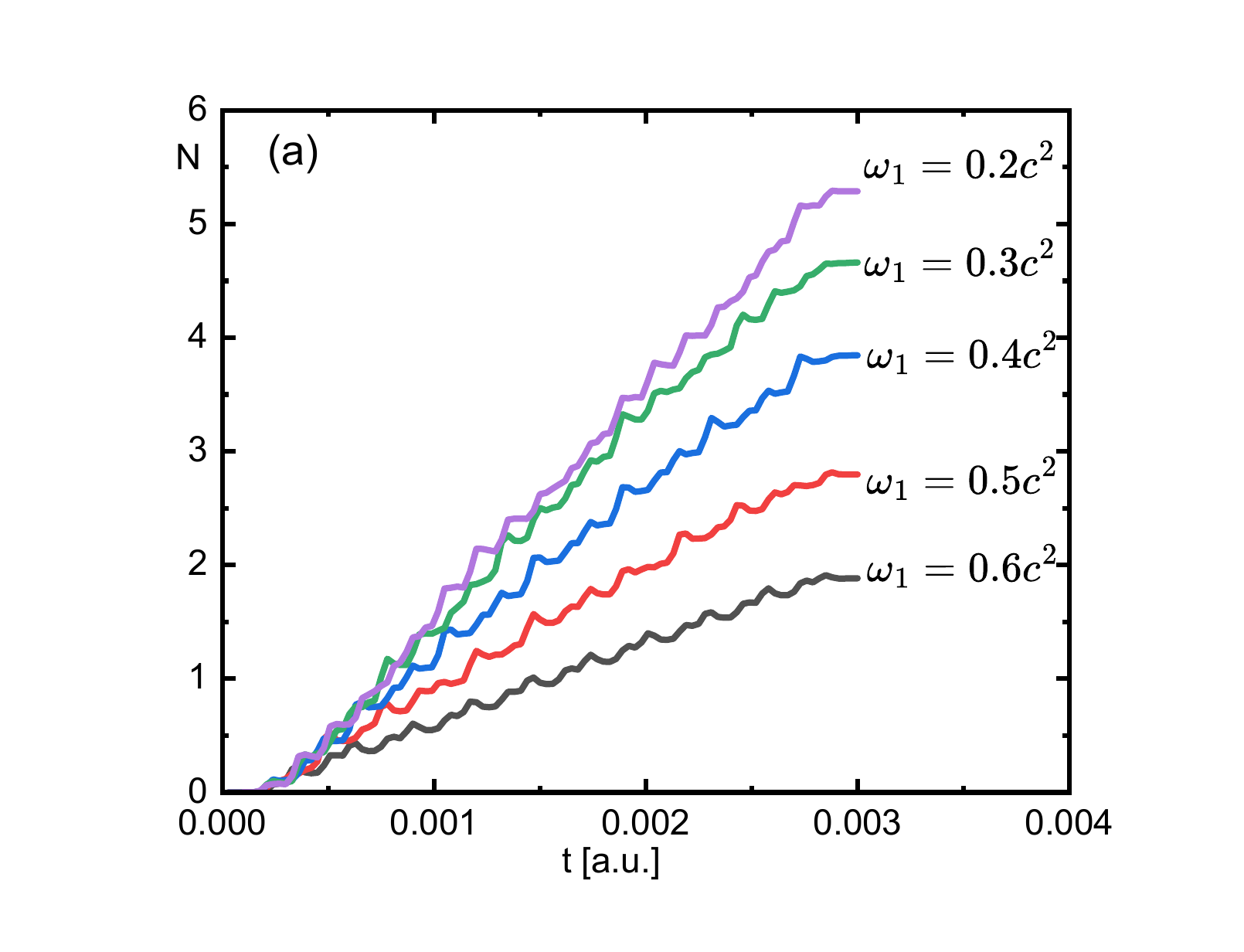}
  \includegraphics[width=0.49\linewidth]{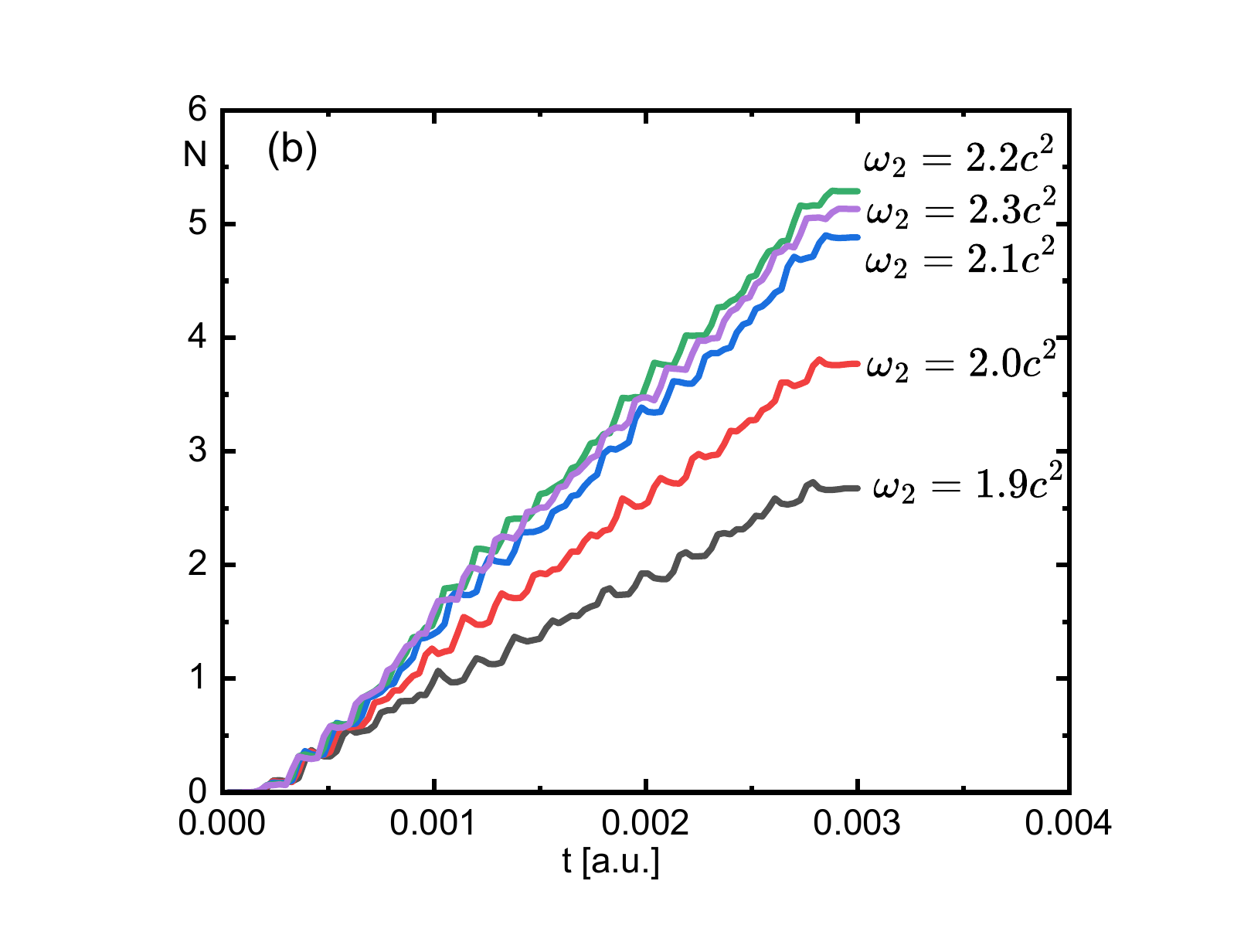}
  \caption{The total pair production as a function of time in a.u., (a) for $\omega_1$ decreases from $0.6c^2$ to $0.2c^2$ while the $\omega_2$ increases from $1.7c^2$ to $2.2c^2$, but the sum of the two frequencies keep constant where $\omega_1+\omega_2=2.4c^2$. (b) for $\omega_1=0.2c^2$ and $\omega_2$ varies from $1.9c^2$ to $2.3c^2$. With the potential well width of $D_1=D_2=20/c$, and other parameters are same with the Fig.~\ref{fig:4}.}
  \label{fig:5}
\end{figure*}

  To observe this assumption, we have investigated the yield of the pair production with different frequencies $\omega_1$ and $\omega_2$, but the sum of them keep constant $\omega_1+\omega_2=2.4c^2$ in Fig.~\ref{fig:5}(a). As a result, the set $\omega_1=0.2c^2, \omega_2=2.2c^2$ has the highest pair production yield. To explain the sensitive of small frequency to the production rate, we have also demonstrated the yield for $\omega_1=0.2c^2$ and $\omega_2$ has been varied from $1.9c^2$ to $2.3c^2$ in Fig.~\ref{fig:5}(b).

  As shown in Fig.~\ref{fig:5}(a), the yield of pair production increases as the $\omega_1$ starts to decrease, but the same time $\omega_2$ increase due to the sum of the two frequencies keep constant in the overlapped fields. The influence of higher frequency $\omega_2$ is vital, but the role of smaller frequency $\omega_1$ can not be ignored. In Fig.~\ref{fig:5}(b), we set the $\omega_1=0.2c^2$ and varies the $\omega_2$ from $1.9c^2$ to $2.3c^2$. Clearly, the $\omega_1=0.2c^2, \omega_2=2.2c^2$ set has the highest pair production yield even higher than the $\omega_1=0.2c^2, \omega_2=2.3c^2$, see Fig.~\ref{fig:5}(b). Generally, $\omega_1=0.2c^2$ and $\omega_2=2.2c^2$ are the best fit for frequency selection to create highest yield of pair production in our overlapped fields due to the lowest frequency $\omega_1$ plays an important role as a dynamically assisted field.

  \begin{figure*}[!htb]
    \includegraphics[width=0.49\linewidth]{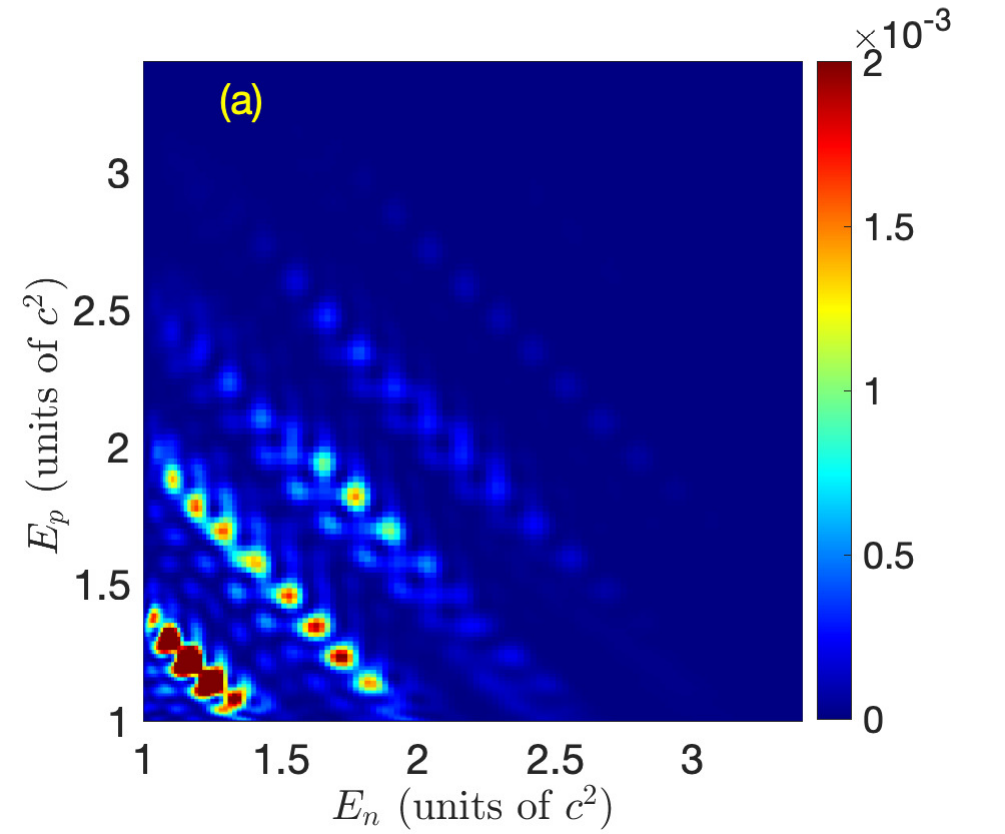}
    \includegraphics[width=0.49\linewidth]{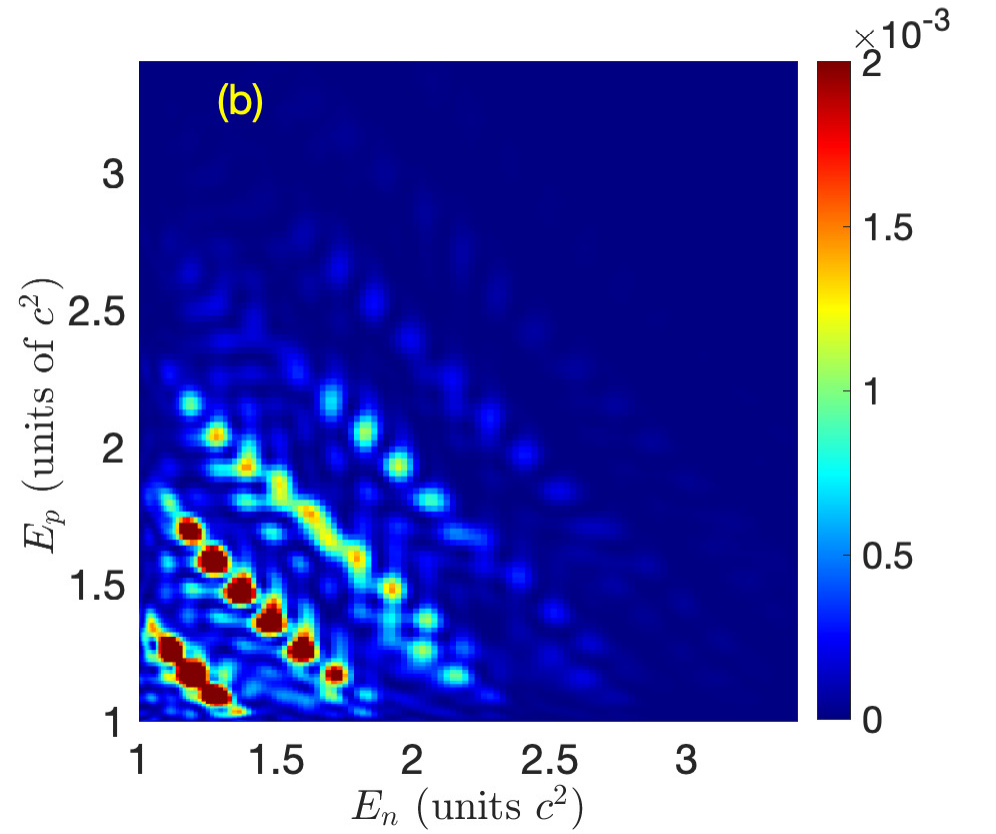}
    \includegraphics[width=0.49\linewidth]{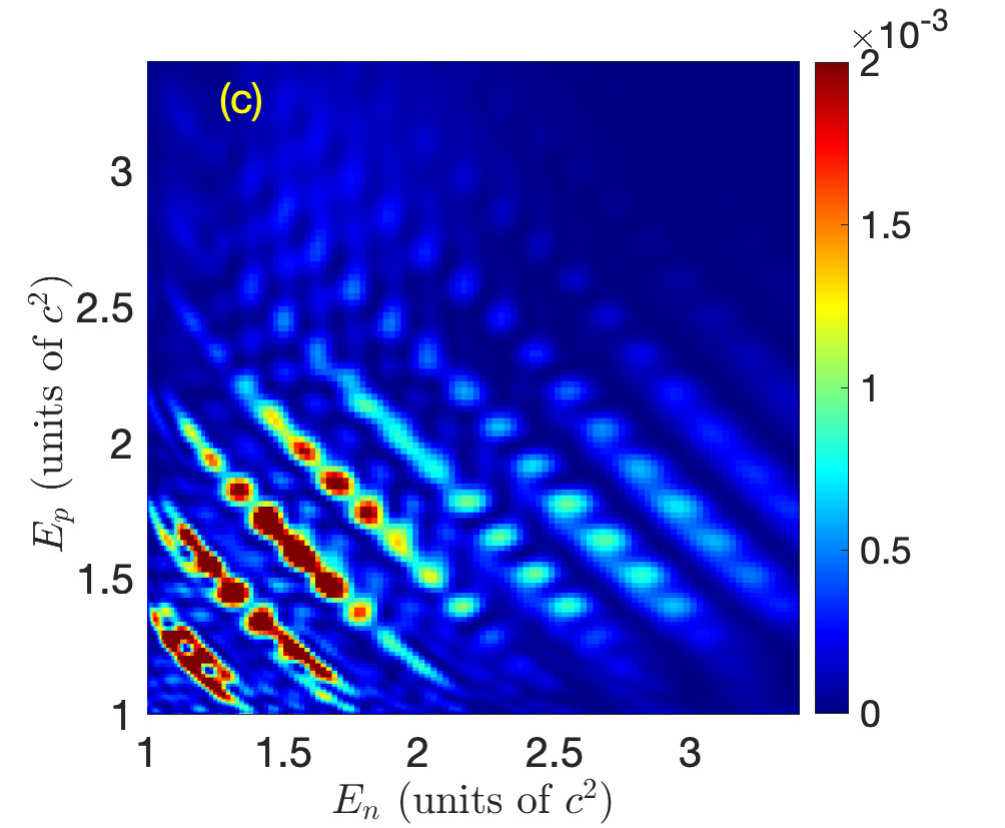}
    \includegraphics[width=0.49\linewidth]{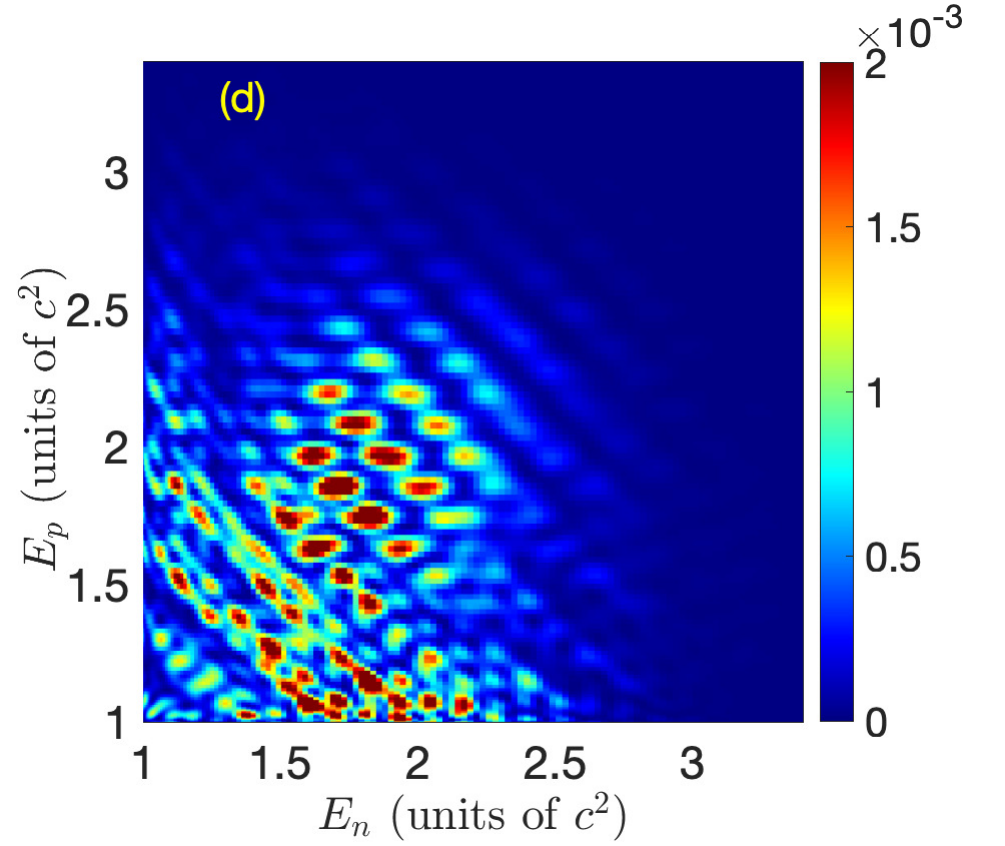}
  \includegraphics[width=0.49\linewidth]{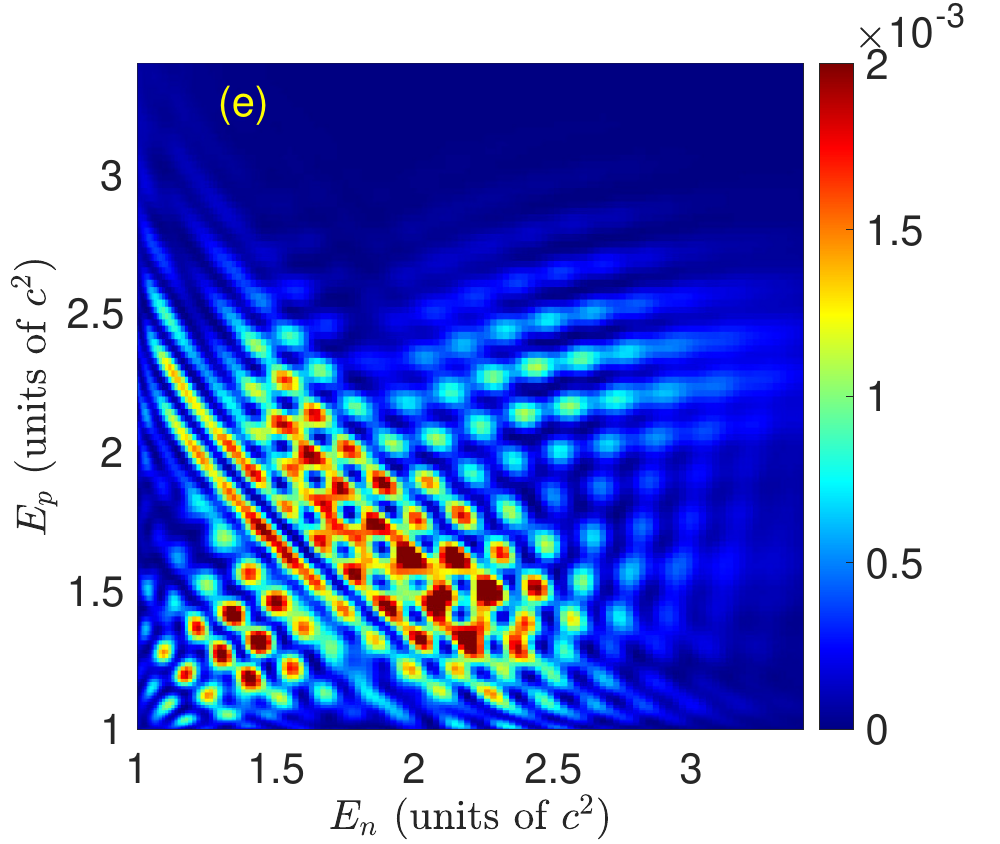}
  \includegraphics[width=0.47\linewidth]{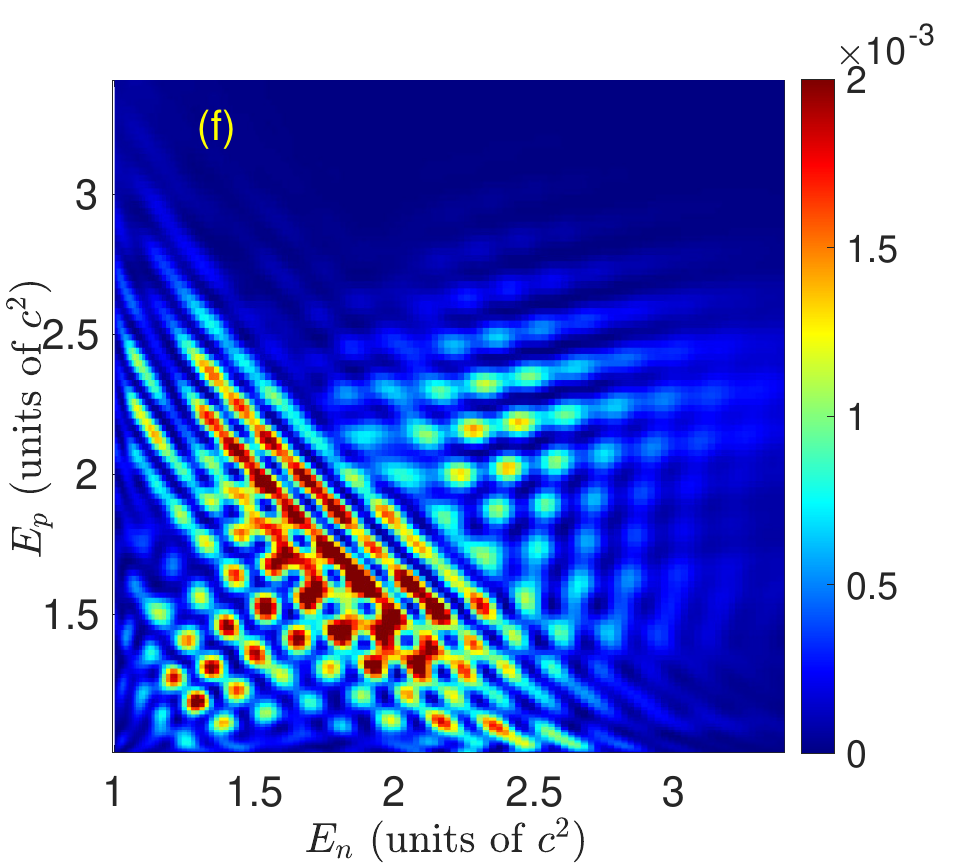}
    \caption{Contour plot of probability distribution of created pairs transition energy corresponding to the Fig.~\ref{fig:5}. From (a) to (e), the frequency $\omega_1$ varies from $0.6c^2$ to $0.2c^2$ and $\omega_2$ varies from $1.8c^2$ to $2.2c^2$, the sum of them keep constant $\omega_1+\omega_2=2.4c^2$, but (f) is for $\omega_1=0.2c^2$ and $\omega_2=2.3c^2$. Other parameters are same with the Fig.~\ref{Fig:Data1}.}
    \label{fig:6}
  \end{figure*}

  In order to get the result for further study, we have presented the corresponding transition energy distribution probabilities $U_{pn}$ for those sets shown in Fig.~\ref{fig:5}. We can be able to differ the variation of the pair production processes through $U_{pn}$ conveniently, see Fig.~\ref{fig:6}.

Now we can compare these two figures Fig.~\ref{fig:5}(a) and Fig.~\ref{fig:6} to get more precise information that the yield of pair production in different frequencies can be affected by some processes and leads to significant enhancement. There have emerged new processes to create more electron-positron pairs as the $\omega_1$ decreases from $0.6c^2$ to $0.2c^2$ and $\omega_2$ increases from $1.8c^2$ to $2.2c^2$ at the same time in order to keep $\omega_1+\omega_2$ constant as $2.4c^2$. Obviously, this results are conform to the yield of the pair production which is also increased exactly in same condition. In Fig.~\ref{fig:6}(a) where $\omega_1=0.6c^2, \omega_2=1.8c^2$ and (b) where $\omega_1=0.5c^2, \omega_2=1.9c^2$, there are respectively three ($\omega_1+\omega_2, 2\omega_1+\omega_2, 3\omega_1+\omega_2$) and four ($\omega_1+\omega_2, 2\omega_1+\omega_2, 3\omega_1+\omega_2,4\omega_1+\omega_2$) main processes  are responsible for the creation of pairs. However, there have emerged many new processes after $\omega_1=0.4c^2, \omega_2=2.0c^2$ shown in Fig.~\ref{fig:6}(c) and more than four-photon processes have started to be very clear to be observed, even have become the dominant processes, see the Fig.~\ref{fig:6}(d)(e)(f).

As for the reason why the set of $\omega_1=0.2c^2$ and $\omega_2=2.2c^2$ has the higher pair production yield than the set of $\omega_1=0.2c^2$ and $\omega_2=2.3c^2$, we have observed that there is slight optimization for $\omega_1=0.2c^2$ and $\omega_2=2.2c^2$ in $\omega_1+\omega_2, 2\omega_1+\omega_2, 3\omega_1+\omega_2$ processes after comparing the $U_{pn}$ result in Fig.~\ref{fig:6}(e) and (f).

  \section{Summary}\label{sec:4}
  The details regarding the influence of potential well width on electron-positron pair production have been thoroughly examined in this study. Analytical and mathematical approaches within the CQFT framework were employed to obtain numerical solutions to the Dirac equation across all time and space. The investigation delves into variations in production yield during the multiphoton pair production process, with a specific focus on dominant processes that involve the absorption of an integer number of photons.

  Notably, distinct changes in specific multiphoton absorption processes were observed as the potential well width expanded, with particular attention given to the $3\omega_1+\omega_2$ process, which exhibited noteworthy effects. The yield of pair production has increased linearly as the $D_1=D_2$ increase to $9/c$ and starts to varied periodically. We have explained this phenomenon by solving the first level perturbative expression. The highest yield of pair production with condition $D_1=D_2$ has been explained by the exact overlapped of the two oscillated fields. As the width of potential well increase, the yield of pair production reaches the highest value at $D_1=D_2$ and they have same value of yield when $|D_1-D_2|\geq5/c$ for different $D_1$ and $D_2$ sets.

  Also, there is slight optimization in the yield of pair production for the smaller frequency $\omega_1$ when $D_1$ and $D_2$ change separately in Fig.~\ref{Fig:Data1}(a) (red line and blue line). We have investigated the pair production yield with different set of frequencies $\omega_1$ and $\omega_2$ in order to study the influence of the smaller frequency. In result, the higher frequency is not only the main reason for the enhancement of the pair production, but the lower frequency of the oscillated field is also playing an important role in pair production process. The set $\omega_1=0.2c^2$ and $\omega_2=2.2c^2$ has the highest pair production yield comparing to the other sets due to the effective of small number photons absorption in electron-positron pair production process.

  \section{Acknowledgments}\label{sec:8}

Some helpful discussions with A. Orkash and C.K. Li are acknowledged. This work was supported by the National Natural Science Foundation of China(NSFC) under Grants No. 11974419, No. 12375240, No.11935008 and the Strategic Priority Research Program of Chinese Academy of Sciences under Grant No. XDA25051000.

\end{document}